\documentclass{cpbtex}

\usepackage{graphicx}
\usepackage{bm}
\usepackage{placeins}

\newcommand{\twobytwofig}{10cm}

\begin{document}
\hypersetup{
  hidelinks,
  pdfborder={0 0 0}
}
\begin{CJK*}{GBK}{song}

\title{Third-order transitions in Ising and Potts models on Watts--Strogatz small-world networks}

\author{\protect\parbox{0.92\textwidth}{\centering
Fangfang Wang$^{1,5,6}$, Wei Liu$^{3}$, Ke Zhang$^{1,5,6}$, Yongjian He$^{1,5,6}$, Kai Qi$^{4}$, Ying Tang$^{2}$, and Zengru Di$^{1,5,6}$\\[4pt]
$^{1}$Department of Systems Science, Faculty of Arts and Sciences, Beijing Normal University, Zhuhai 519087, China\\
$^{2}$Institute of Fundamental and Frontier Sciences, University of Electronic Science and Technology of China, Chengdu 611731, China\\
$^{3}$College of Science, Xi'an University of Science and Technology, Xi'an, China\\
$^{4}$2020 X-Lab, Shanghai Institute of Microsystem and Information Technology, Chinese Academy of Sciences, Shanghai, China\\
$^{5}$International Academic Center of Complex Systems, Beijing Normal University, Zhuhai, China\\
$^{6}$School of Systems Science, Beijing Normal University, Beijing 100875, China\\[4pt]
\small Corresponding authors: zdi@bnu.edu.cn; jamestang23@gmail.com; weiliu@xust.edu.cn
}}

\date{}
\maketitle

\begin{abstract}
We study third-order transitions in the two-dimensional Ising and Potts model on regular lattices and Watts--Strogatz small-world networks. 
Cluster observables are used to track post-critical boundary reorganization and pre-critical cluster breakup. 
For the Ising model, the critical temperature $T_c$ is calibrated independently from Binder-cumulant crossings and susceptibility peaks, whereas for the Potts model on small-world networks it is identified operationally from the dominant critical peak of $\mathrm d\langle P\rangle/\mathrm dT$. The independent and dependent third-order transitions are identified from the isolated-spin peak and the post-critical structural extremum, respectively. 
For both lattice and small-world topologies, we find the robust ordering $T_{\mathrm{ind}}<T_c<T_{\mathrm{dep}}$. 
Increasing the rewiring probability shifts all three characteristic temperatures upward and enhances the visibility of the post-critical transition. 
The effect is especially clear in the Potts model, where perimeter-based observables are more sensitive to multistate boundary fluctuations. 
The systematic persistence of the characteristic temperature hierarchy across topologies and finite sizes argues against interpreting these features as incidental finite-size irregularities. 
Instead, our results support their interpretation as genuine third-order transitions whose structural detectability can be amplified by network topology.
\end{abstract}


\section{Introduction}
\label{sec:intro}

Phase transitions are traditionally characterized by nonanalytic behavior of thermodynamic functions \cite{Landau1999PhaseTransitionsCiSE}. 
In the Ehrenfest sense, a first-order transition is associated with a singularity in the first derivative of the free energy, and a second-order transition with a singularity in the second derivative while lower derivatives remain continuous \cite{Ehrenfest1933Amsterdam}. 
In finite systems, however, these singularities are rounded, and the transition region can contain additional reorganizations beyond the dominant low-order critical feature \cite{Fisher1972FiniteSizeScalingPRL}. 
Such higher-order behavior is physically important because it reflects subtle changes in domains, boundaries, and mesoscopic correlations that are not fully captured by conventional thermodynamic observables alone \cite{Gross2001MicrocanonicalThermodynamics,Qi2018MIPAClassification}. 
At the same time, these transitions are much harder to resolve numerically than ordinary first- or second-order transitions, because their signatures are weaker and can be obscured by finite-size rounding and background critical fluctuations.

A central advance in this field is the establishment of third-order transitions as a legitimate object of systematic study, rather than as incidental finite-size irregularities or purely formal higher-derivative curiosities \cite{Wang2026Canonical}. 
A major route toward this understanding has been microcanonical inflection-point analysis (MIPA), which identifies higher-order transitions from derivatives of the microcanonical entropy together with the principle of minimal sensitivity \cite{Stevenson1981OPT,Stevenson1981RSA}. 
MIPA has revealed higher-order transitions in a broad class of finite systems, including spin models, polymers, and related statistical systems \cite{liu2022pseudo,wangff,BelHadjAissa2021KT,Aierken2023,Aierken2023Bifurcation,Corti2023,DiCairano2024Topological,DiCairano2022Topological,Rocha2025Microcanonical,Junghans2006PeptideAggregation,Schnabel2011EntropyInflection}. 
In parallel, canonical real-space studies have shown that cluster observables can provide an intuitive structural characterization of third-order transitions. 
For the two-dimensional Ising model, the low-temperature proliferation of isolated spins and the post-critical acceleration of cluster reorganization furnish natural markers of independent and dependent third-order transitions, respectively \cite{sitarachu2022evidence,Sitarachu2020ThirdOrderIsing}. 
These results show that third-order transitions are not merely formal higher-order curiosities, but can appear as observable manifestations of cluster breakup, defect seeding, and boundary reshaping in finite spin systems \cite{shi2026pseudo,liu2025PLA,Liu2024MV}.

A key unresolved question is whether such third-order transitions remain robust under topological heterogeneity. 
Most existing structural studies of third-order transitions have focused on regular lattices, where geometry is fixed and local connectivity is uniform \cite{sitarachu2022evidence,shi2026pseudo,liu2025PLA,Liu2024MV}. 
By contrast, many interacting systems are embedded on heterogeneous or shortcut-rich networks, where long-range links can modify fluctuation propagation, cooperative reorganization, and the visibility of structural transition markers \cite{AlbertBarabasi2002RMP,Dorogovtsev2008CriticalPhenomenaNetworks}. 
Among network models, Watts--Strogatz small-world networks are especially suitable for addressing this problem because a single rewiring probability continuously interpolates between a regular lattice and a topology with increasing shortcut density \cite{SWN}, while also providing a well-established setting for studying topology-modified critical behavior in spin systems \cite{Gitterman2000SmallWorldIsing,Herrero2002IsingSmallWorld,Herrero2004FerroSmallWorld}. 
This makes them a controlled platform for testing whether the characteristic temperature hierarchy of third-order transitions survives topological disorder, and whether shortcut-enhanced connectivity suppresses, preserves, or even amplifies their structural detectability. 
While conventional critical behavior on complex networks has been widely studied \cite{AlbertBarabasi2002RMP}, the topology dependence of third-order transitions remains comparatively unexplored.

In this work, we investigate third-order transitions in the two-dimensional Ising and three-state Potts models on regular lattices and Watts--Strogatz small-world networks. 
We first determine the conventional critical temperature $T_c$ by model-appropriate criteria: Binder-cumulant crossings and susceptibility peaks for the Ising model, and the dominant critical peak of $\mathrm d\langle P\rangle/\mathrm dT$ for the Potts model on small-world networks. We then identify the independent and dependent third-order transitions from cluster-based structural observables. 
Our goal is not a precision study of critical exponents, but a structural characterization of how the transition hierarchy and its visibility evolve with topology. 
We show that both models exhibit the robust ordering $T_{\mathrm{ind}}<T_c<T_{\mathrm{dep}}$ on both regular lattices and small-world networks. 
Moreover, increasing rewiring shifts all three characteristic temperatures upward and, in particular, enhances the visibility of the post-critical transition. 
The systematic persistence of this hierarchy across the topologies and finite sizes studied argues against interpreting these features as incidental finite-size irregularities. 
Instead, our results support their interpretation as genuine third-order transitions whose structural detectability can be amplified by network topology.

\section{Models and methods}
\label{sec:methods}

\subsection{Watts--Strogatz networks and spin models}
\label{subsec:models}

We start from a two-dimensional $L\times L$ square lattice with periodic boundary conditions and nearest-neighbor connectivity. 
To introduce small-world shortcuts while preserving the total number of edges, each undirected lattice edge is rewired independently with probability $p$. 
For a selected edge $(i,j)$, the endpoint $j$ is replaced by a uniformly sampled node $k\neq i$, subject to the constraints that self-loops and multi-edges are forbidden. 
In this way, $p=0$ recovers the regular square lattice, whereas larger $p$ generates increasingly shortcut-rich Watts--Strogatz (WS) topologies.

On the resulting graph $G=(V,E)$, we consider two ferromagnetic spin models. 
For the Ising model \cite{ising},
\begin{equation}
  \mathcal{H}_\mathrm{Ising} = -J \sum_{\langle ij\rangle\in E} s_i s_j,\qquad s_i\in\{\pm1\},
  \label{eq:ising_hamiltonian}
\end{equation}
and for the three-state Potts model \cite{Potts1952,WuRMP1982},
\begin{equation}
  \mathcal{H}_\mathrm{Potts} = -J \sum_{\langle ij\rangle\in E} \delta(\sigma_i,\sigma_j),\qquad \sigma_i\in\{0,1,2\}.
  \label{eq:potts_hamiltonian}
\end{equation}
Here $J>0$ is the ferromagnetic coupling, $\delta(\cdot,\cdot)$ is the Kronecker delta, and Boltzmann's constant is set to $k_B=1$ throughout.

Unless otherwise stated, the system size $L$ and rewiring probability $p$ are specified in the corresponding simulations and figures.
\begin{equation}
L\in\{30,42,48,60,90,180,360,480\},\qquad
p\in\{0,0.1,0.3,1.0\},
\label{eq:L_p_range}
\end{equation}
where $p=0$ denotes the regular square lattice.

\subsection{Monte Carlo sampling}
\label{subsec:mc}

Configurations are generated with the Swendsen--Wang cluster algorithm \cite{Ferrenberg1988MonteCarlo}. 
For the Ising model, a bond between like-spin neighbors is activated with probability
\begin{equation}
  p_\mathrm{add}=1-e^{-2\beta J},
  \label{eq:ising_padd}
\end{equation}
where $\beta=1/T$. 
For the three-state Potts model, the activation probability is
\begin{equation}
  p_\mathrm{add}=1-e^{-\beta J}.
  \label{eq:potts_padd}
\end{equation}
The activated bonds define Fortuin--Kasteleyn clusters \cite{Swendsen1987Nonuniversal}. 
For the Ising model, each cluster is flipped with probability $1/2$; for the Potts model, each cluster is assigned a new spin value drawn uniformly from the three allowed states. 
One Monte Carlo step (MCS) consists of one full bond-activation and cluster-update sweep.

For the Ising model, observables are sampled on a temperature grid of 30 linearly spaced points covering the interval $T\in[1,4]$, corresponding to a step size $\Delta T=3/29\approx0.10345$. 
For the three-state Potts model, observables are sampled on a temperature grid of 30 linearly spaced points covering the interval $T\in[0.5,1.5]$, corresponding to a step size $\Delta T=1/29\approx0.03448$. 
At each temperature, $10^{5}$ Monte Carlo steps are performed, with the first $5\times10^{4}$ steps discarded for equilibration and the remaining $5\times10^{4}$ steps used for measurements. 
For the small-world simulations, the network is generated with degree $k=4$ and rewiring probabilities $p=0.1$, $0.3$, and $1.0$. For each chosen system size and rewiring probability, a Watts--Strogatz network is first generated and then kept fixed throughout the temperature sweep as a quenched topology. Temperature derivatives such as $\mathrm d\langle P\rangle/\mathrm dT$ and $\mathrm dW/\mathrm dT$ are evaluated numerically from the Monte Carlo data on the discrete temperature grid, and the characteristic temperatures are located directly from the corresponding extrema in the sampled curves.

From the sampled temperature series, we evaluate the temperature derivatives of the structural observables numerically and locate the relevant extrema on the size-resolved curves. In the Ising model, $T_c$ is determined independently from Binder-cumulant crossings and susceptibility maxima, whereas $T_{\mathrm{ind}}$ and $T_{\mathrm{dep}}$ are extracted from structural observables. In the Potts model on small-world networks, the dominant peak of $\mathrm d\langle P\rangle/\mathrm dT$ is used as the operational estimator of $T_c$, and the subsequent post-critical extremum of the same curve defines $T_{\mathrm{dep}}$.

\subsection{Cluster observables and characteristic temperatures}
\label{subsec:cluster}

\begin{figure}[htbp]
\centering
\includegraphics[width=5.5cm]{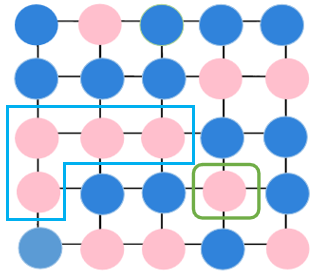}
\caption{Illustration of the cluster observables used in the analysis. The cluster size $A(C)$ counts the number of sites in a connected equal-spin cluster, the perimeter $P(C)$ counts the number of boundary edges, and $n_{\mathrm{iso}}$ counts isolated spins.}
\label{fig:cluster_obs}
\end{figure}

A cluster $C\subseteq V$ is defined as a connected component of equal-spin sites \cite{sitarachu2022evidence,Liu2025PottsGeometry}. 
For each cluster, we define its size and perimeter as
\begin{equation}
  A(C)=|C|,
  \label{eq:cluster_area}
\end{equation}
\begin{equation}
  P(C)=\big|\{(i,j)\in E:\ i\in C,\ j\notin C\}\big|.
  \label{eq:cluster_perimeter}
\end{equation}
Thus, $A(C)$ counts the number of sites in the cluster, whereas $P(C)$ counts the number of boundary edges connecting the cluster to its exterior. Fig.~\ref{fig:cluster_obs} illustrates these geometric definitions together with the isolated-spin observable $n_{\mathrm{iso}}$ used below.

To exclude isolated single-site objects when evaluating typical cluster geometry, we define
\begin{equation}
  \mathcal{C}_2=\{C:\ A(C)\ge 2\},
  \label{eq:C2}
\end{equation}
with $n_c=|\mathcal{C}_2|$. 
The average cluster size and average cluster perimeter are then
\begin{eqnarray}
  \langle A\rangle &=& \frac{1}{n_c}\sum_{C\in\mathcal{C}_2} A(C),\label{eq:avgA}\\
  \langle P\rangle &=& \frac{1}{n_c}\sum_{C\in\mathcal{C}_2} P(C).\label{eq:avgP}
\end{eqnarray}

We also monitor the largest-cluster size $A_{\max}$ and the number of isolated spins. 
Introducing a generic local state variable $x_i$ ($x_i=s_i$ for Ising and $x_i=\sigma_i$ for Potts), the number of isolated spins is defined as
\begin{equation}
  n_{\mathrm{iso}}
  =
  \big|
  \{\,i\in V:\ \nexists\, j\in N(i)\ \text{such that}\ x_j=x_i\,\}
  \big|,
  \label{eq:niso}
\end{equation}
where $N(i)$ is the neighbor set of node $i$.

In the regular-lattice benchmark, the Ising dependent third-order transition is most visible in area-based observables, whereas in the Potts model and on shortcut-rich small-world networks the perimeter derivative is more sensitive to boundary reorganization. In the main text we therefore adopt the following operational definitions. For the Ising model, $T_c$ is determined independently from Binder-cumulant crossings and susceptibility peaks. For the Potts model on small-world networks, $T_c$ is taken as the dominant critical peak of $\mathrm d\langle P\rangle/\mathrm dT$. In both models, $T_{\mathrm{ind}}$ is determined from the maximum of $\langle n_{\mathrm{iso}}\rangle$, and $T_{\mathrm{dep}}$ from the post-critical local extremum of the structural derivative, taken in the small-world analysis as the extremum of $\mathrm d\langle P\rangle/\mathrm dT$.

For completeness, we also examined alternative structural indicators, including the largest-cluster observable and the auxiliary Potts quantity
\begin{equation}
  W=\frac{1}{n_c}\sum_{C\in\mathcal{C}_2} A(C)\left[\frac{P(C)}{z\,A(C)}\right]^\alpha,
  \label{eq:W}
\end{equation}
where $z=4$ and $0<\alpha\le 1$. 
These alternative indicators are used only as supplementary consistency checks rather than as the primary definition of characteristic temperatures.

\subsection{Finite-size analysis of the critical temperature}
\label{subsec:fss}

To calibrate the conventional critical temperature on small-world networks for the Ising model, we measure the absolute magnetization,
\begin{equation}
m_{\rm abs}=\Big\langle \Big| \frac{1}{N}\sum_{i\in V}s_i \Big| \Big\rangle,
\label{eq:mabs}
\end{equation}
the magnetization
\begin{equation}
m_0=\frac{1}{N}\sum_{i\in V}s_i,
\label{eq:m0}
\end{equation}
the susceptibility
\begin{equation}
\chi=N\beta\left(\langle m_0^2\rangle-\langle m_0\rangle^2\right),
\label{eq:chi}
\end{equation}
and the fourth-order Binder cumulant
\begin{equation}
U_4=1-\frac{\langle m_0^4\rangle}{3\langle m_0^2\rangle^2}.
\label{eq:binder}
\end{equation}
Here $N=|V|$ is the number of nodes.

In the main text, finite-size analysis is used only to calibrate $T_c$ for the Ising model independently from the cluster-based observables. Specifically, for the Ising model we determine $T_c$ from Binder-cumulant crossings and from the size dependence of the susceptibility peak position. By contrast, in the Potts model on small-world networks the dominant peak of $\mathrm d\langle P\rangle/\mathrm dT$ is used as the operational critical-point estimator, and the subsequent post-critical extremum of the same curve identifies $T_{\mathrm{dep}}$. A detailed precision analysis of critical exponents is not the focus of the present work. Additional supporting discussion, auxiliary structural observables, and summary plots are provided in Appendices~A--C.

\section{Results and discussion}
\label{sec:results}

\subsection{Benchmark on regular lattices}
\label{subsec:benchmark}

\subsubsection{Dependent third-order transitions}
\label{subsubsec:benchmark_dep}

\begin{figure}[htbp]
\centering
\includegraphics[width=\twobytwofig]{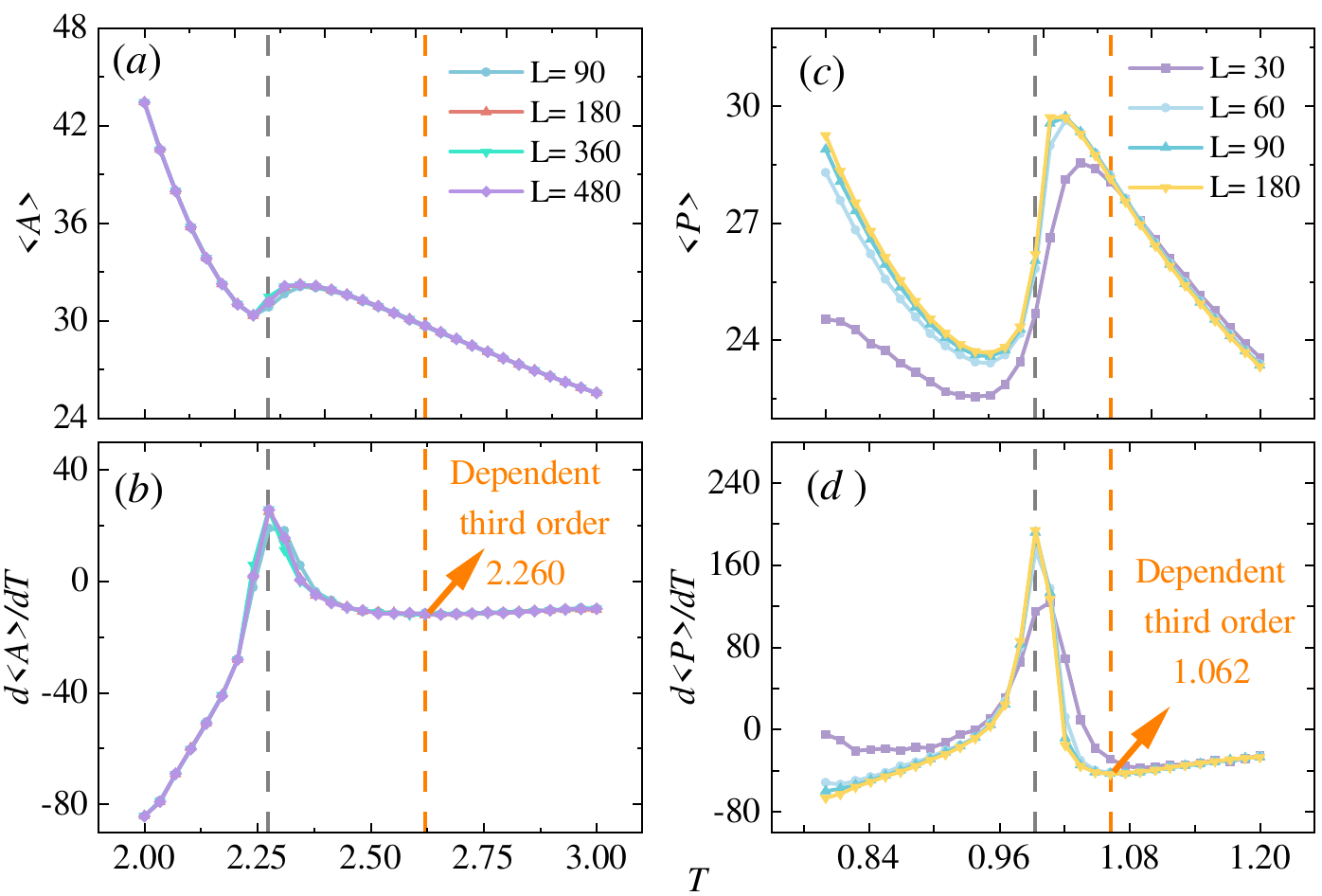}
\caption{Dependent third-order transitions on the regular square lattice. The figure compares the structurally most informative post-critical observables in the Ising and three-state Potts models. In the Ising model, the dependent transition is more clearly resolved in the area-based derivative, whereas in the Potts model it is more clearly resolved in the perimeter-based derivative. Gray dashed lines mark the critical temperature $T_c$, and pink dashed lines indicate the post-critical temperature $T_{\mathrm{dep}}$.}
\label{fig:benchmark_dep}
\end{figure}

We begin with the regular square lattice as a benchmark in order to identify which structural observables are most sensitive to post-critical reorganization in the two models. As shown in Fig.~\ref{fig:benchmark_dep}, the Ising model exhibits the clearest dependent third-order signature in the area-based observable: after the dominant critical feature at $T_c$, the derivative of the average cluster size displays a secondary post-critical extremum, indicating an additional regime of accelerated cluster reorganization on the disordered side. By contrast, the perimeter derivative is dominated by the critical peak and is less effective at separating the post-critical feature on the square lattice. This difference suggests that in the Ising case the dependent transition is more directly tied to the redistribution of cluster mass than to boundary complexity alone.

The Potts model shows a different pattern. In Fig.~\ref{fig:benchmark_dep}, $\mathrm d\langle P\rangle/\mathrm dT$ displays a clear critical peak together with a resolvable post-critical extremum, whereas the area derivative varies more smoothly and does not isolate the dependent feature as clearly. This contrast is physically intuitive: the Ising system is dominated by two-state domain growth and breakup, while the three-state Potts model supports richer boundary rearrangements. These benchmark results motivate our later choice of perimeter-based diagnostics as the primary dependent-transition indicator on small-world networks.

\begin{figure}[htbp]
\centering
\includegraphics[width=\twobytwofig]{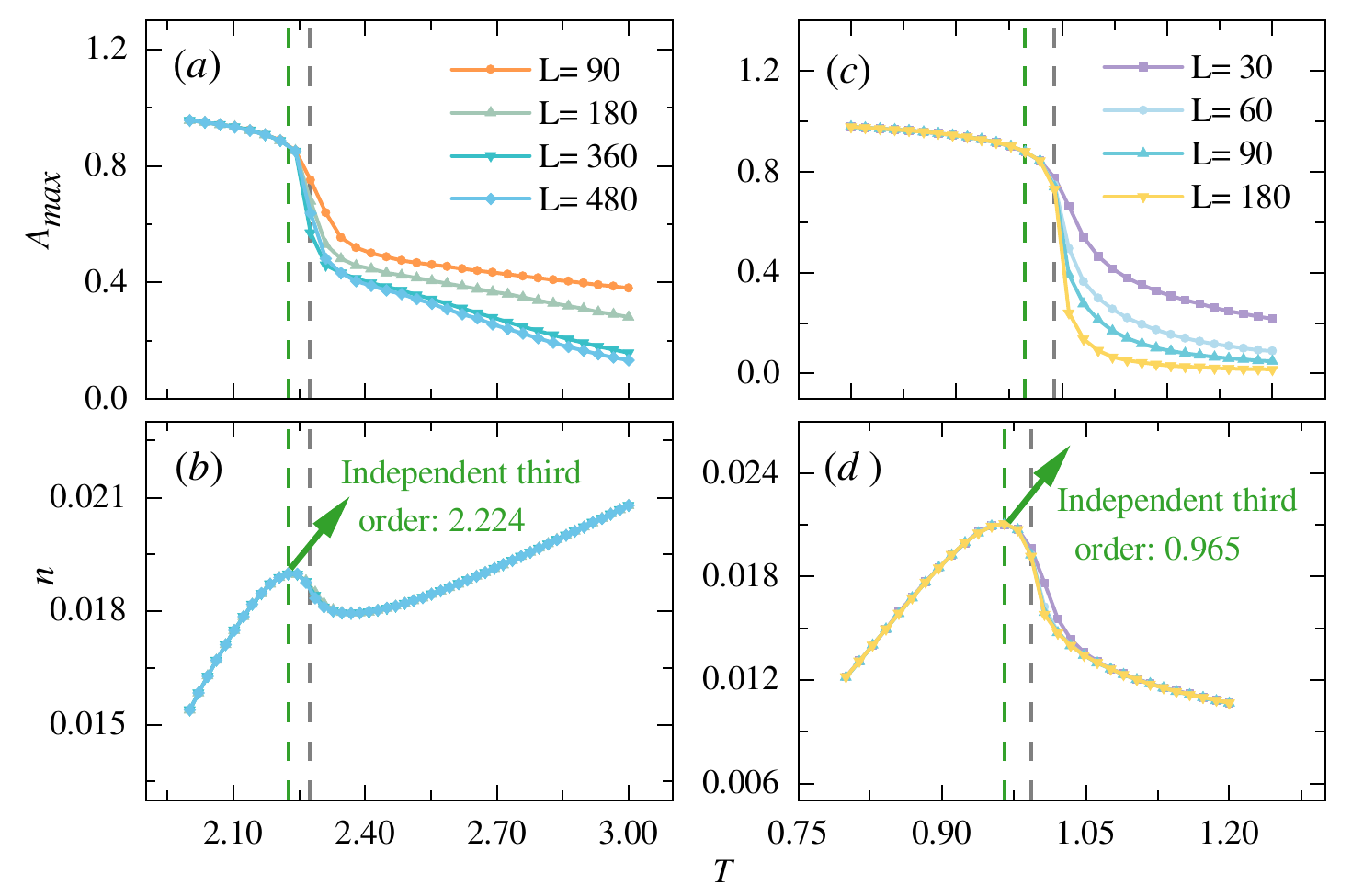}
\caption{Independent third-order transition on the regular square lattice. (a),(c) Largest-cluster size $A_{\max}$ for the Ising and three-state Potts models. (b),(d) Number of isolated spins $n_{\mathrm{iso}}$ for the Ising and three-state Potts models. In both models, the low-temperature-side precursor is most clearly identified by the maximum of $\langle n_{\mathrm{iso}}\rangle$, while $A_{\max}$ provides supporting information on the breakup of large domains. Gray dashed lines mark the critical temperature $T_c$, blue dashed lines mark the independent third-order temperature $T_{\mathrm{ind}}$, and pink dashed lines indicate the post-critical temperature $T_{\mathrm{dep}}$ for reference.}
\label{fig:benchmark_ind}
\end{figure}

We next turn to the independent third-order transition, which occurs on the low-temperature side of the conventional critical point and is associated with the breakup of large clusters and the proliferation of isolated spins. As shown in Fig.~\ref{fig:benchmark_ind}, the number of isolated spins, $n_{\mathrm{iso}}$, provides the cleanest operational indicator of this transition in both the Ising and Potts models. As the temperature approaches $T_c$ from below, large domains begin to fragment and isolated sites become increasingly abundant, causing $\langle n_{\mathrm{iso}}\rangle$ to rise and reach a maximum at $T_{\mathrm{ind}}$.

This behavior is consistent with the simultaneous reduction of the largest-cluster size $A_{\max}$. 
Although $A_{\max}$ itself captures the progressive loss of large connected domains, its characteristic point is less sharply defined than the peak of $\langle n_{\mathrm{iso}}\rangle$. 
For this reason, in the remainder of the paper we use the maximum of $\langle n_{\mathrm{iso}}\rangle$ as the primary operational definition of $T_{\mathrm{ind}}$, while $A_{\max}$ is treated as a supporting structural observable. 
Taken together, the lattice benchmark establishes a simple working picture: the independent transition is most naturally tracked by isolated-spin proliferation, whereas the dependent transition is linked to post-critical structural reorganization and is model dependent in its most sensitive observable.

\subsection{Critical-point calibration on small-world networks}
\label{subsec:calibration}

\begin{figure}[htbp]
\centering
\includegraphics[width=15.0cm]{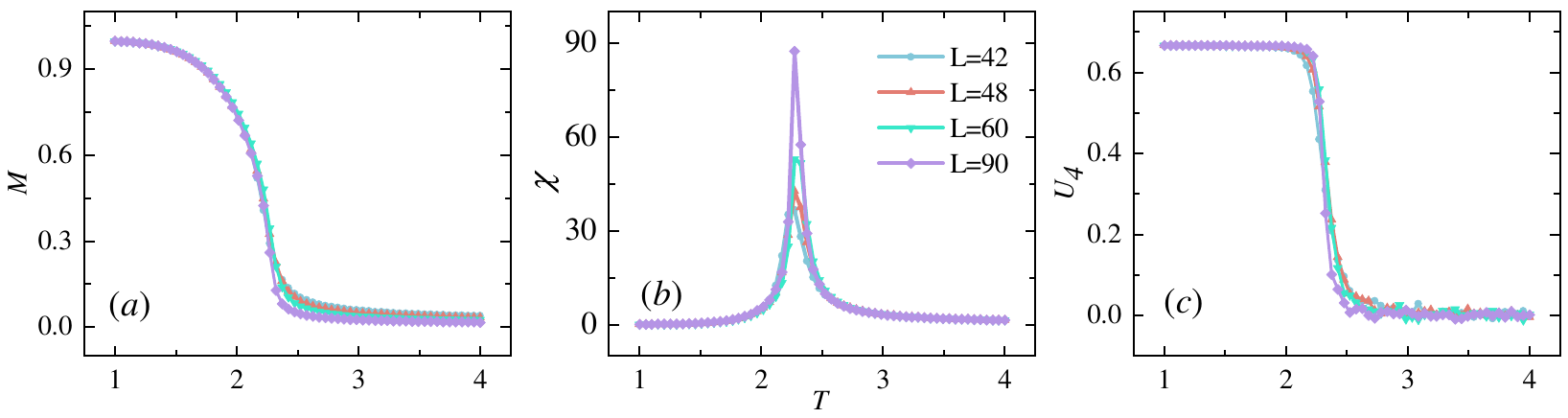}
\caption{Calibration of the conventional critical temperature on Watts--Strogatz small-world networks for the Ising model at rewiring probability $p=0.1$. Panels (a)--(c) show the absolute magnetization, magnetic susceptibility, and fourth-order Binder cumulant. The susceptibility peak and Binder-cumulant crossing provide mutually consistent estimates of $T_c$.}
\label{fig:ising_fss}
\end{figure}

Before discussing higher-order precursor transitions on Watts--Strogatz small-world networks, we first calibrate the conventional critical temperature $T_c$ for the Ising model independently from standard magnetic observables. Fig.~\ref{fig:ising_fss} shows a representative calibration for the Ising model at rewiring probability $p=0.1$. The absolute magnetization decreases more sharply with increasing system size, the susceptibility develops a progressively better defined maximum, and the Binder cumulant exhibits an increasingly coherent crossing. These standard indicators provide a mutually consistent estimate of $T_c$ and therefore separate the identification of the critical point from the later structural analysis in the Ising case.

This independence is essential for the logic of the Ising analysis. The precursor temperatures $T_{\mathrm{ind}}$ and $T_{\mathrm{dep}}$ are meant to describe additional reorganizations on the low- and high-temperature sides of the critical region, not alternative ways of reading $T_c$ from the same structural curve. By calibrating $T_c$ first and only then examining the cluster observables, we reduce the ambiguity that would otherwise arise from multiple extrema in $\langle P\rangle$, $\langle A\rangle$, or their temperature derivatives.
The Potts analysis on small-world networks is operationally different. There, the dominant peak of $\mathrm d\langle P\rangle/\mathrm dT$ is taken as the critical temperature $T_c$, while the subsequent post-critical local extremum of the same curve is used to identify $T_{\mathrm{dep}}$. Because these two features are well separated in temperature, the same perimeter-based derivative can resolve both the main critical reorganization and the dependent precursor.

Because the present work is aimed at establishing topology-dependent precursor structure rather than at extracting precision exponents, the finite-size analysis is used here primarily as a calibration step for the Ising model. The supplementary discussion collected in Appendix~\hyperref[app:A]{A} supports this interpretation but is not used as the central evidence for the main conclusions.

\subsection{Topology dependence of third-order precursor transitions}
\label{subsec:topology}

\subsubsection{Ising model}
\label{subsubsec:ising_sw}

\begin{figure}[htbp]
\centering
\includegraphics[width=15.0cm]{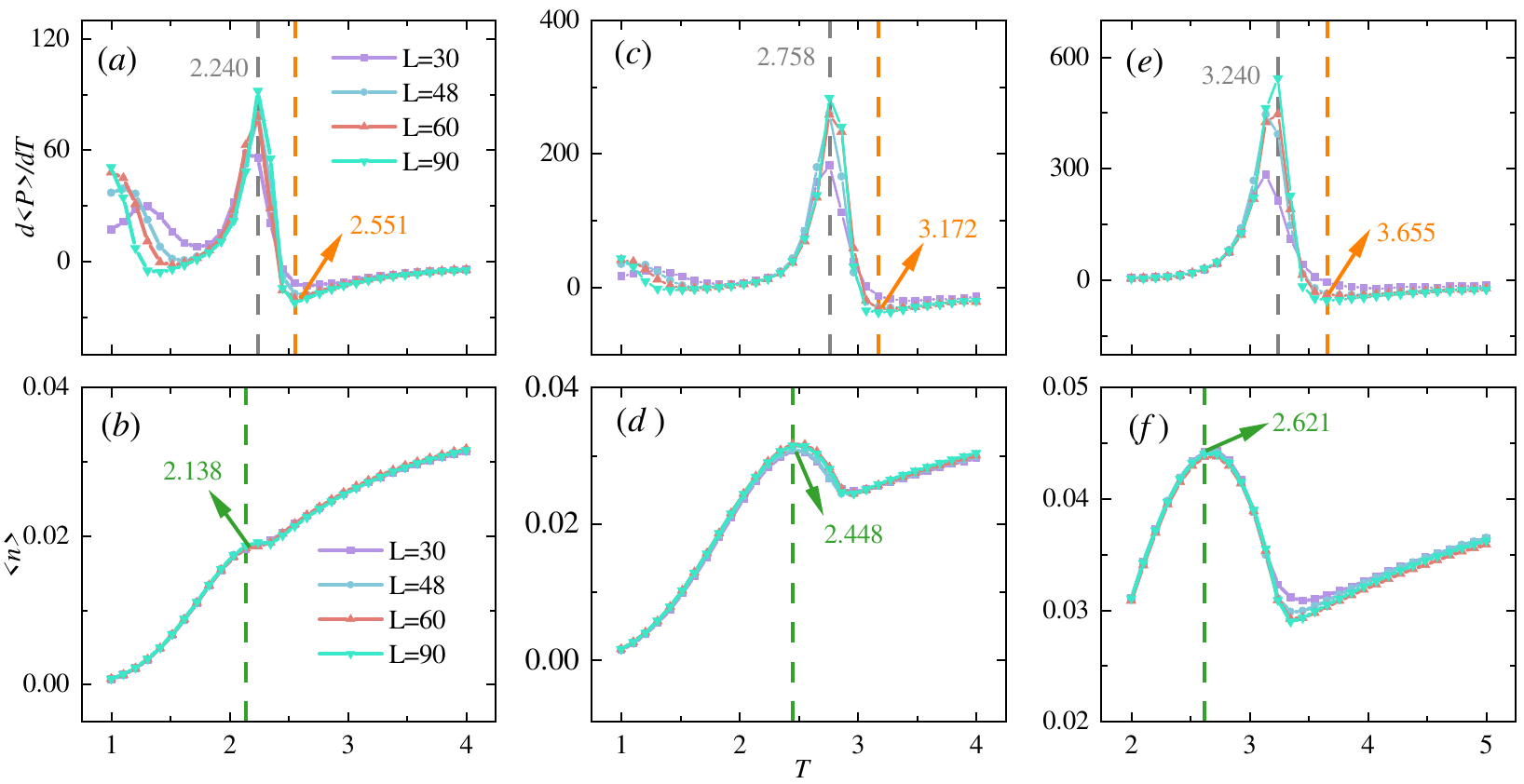}
\caption{Topology dependence of third-order precursor transitions in the Ising model on Watts--Strogatz small-world networks. Panels (a),(b), (c),(d), and (e),(f) correspond to rewiring probabilities $p=0.1$, $0.3$, and $1.0$, respectively. The top row shows the perimeter-based post-critical signal $\mathrm{d}\langle P\rangle/\mathrm{d}T$, and the bottom row shows the isolated-spin indicator $\langle n_{\mathrm{iso}}\rangle$. Increasing rewiring shifts the full hierarchy of characteristic temperatures upward and makes the post-critical structural precursor easier to separate from the dominant critical peak. Gray dashed lines mark $T_c$, blue dashed lines mark $T_{\mathrm{ind}}$, and pink dashed lines mark $T_{\mathrm{dep}}$.}
\label{fig:ising_sw}
\end{figure}

We first examine the Ising model on small-world networks with different rewiring probabilities. In Fig.~\ref{fig:ising_sw}, the left, middle, and right columns correspond to $p=0.1$, $0.3$, and $1.0$, respectively. Relative to the regular lattice benchmark, the introduction of shortcuts shifts the full temperature hierarchy upward and makes the post-critical structural feature easier to resolve. In particular, the post-critical extremum of $\mathrm{d}\langle P\rangle/\mathrm{d}T$ moves from about $T_{\mathrm{dep}}\approx 2.55$ at $p=0.1$ to $\approx 3.17$ at $p=0.3$ and $\approx 3.66$ at $p=1.0$, while the independently calibrated critical temperature shifts from $T_c\approx 2.24$ to $\approx 2.76$ and then to $\approx 3.24$.

Two points are particularly important. First, the post-critical feature becomes more clearly separated from the dominant critical peak as $p$ increases. This indicates that long-range shortcuts amplify the cooperative boundary reshaping that follows the principal critical reorganization. Second, the low-temperature isolated-spin maximum remains well defined across all three rewiring probabilities. Its location shifts more moderately---from about $T_{\mathrm{ind}}\approx 2.14$ at $p=0.1$ to $\approx 2.45$ at $p=0.3$ and $\approx 2.62$ at $p=1.0$---but the qualitative precursor mechanism is unchanged: large ordered domains begin to fragment before the critical point, and the number of isolated spins reaches a maximum at that stage.

Taken together, the Ising results show that shortcut-enhanced connectivity does not wash out the precursor structure. Instead, it preserves the ordering
\begin{equation}
T_{\mathrm{ind}} < T_c < T_{\mathrm{dep}},
\label{eq:hierarchy_ising}
\end{equation}
while simultaneously increasing the temperature scale of all three characteristic features and sharpening the post-critical signal. In the Ising case, the topology effect is therefore strongest in the dependent transition, whose visibility is controlled by the reorganization of cluster boundaries after the main critical event.

\subsubsection{Potts model}
\label{subsubsec:potts_sw}

\begin{figure}[htbp]
\centering
\includegraphics[width=15.0cm]{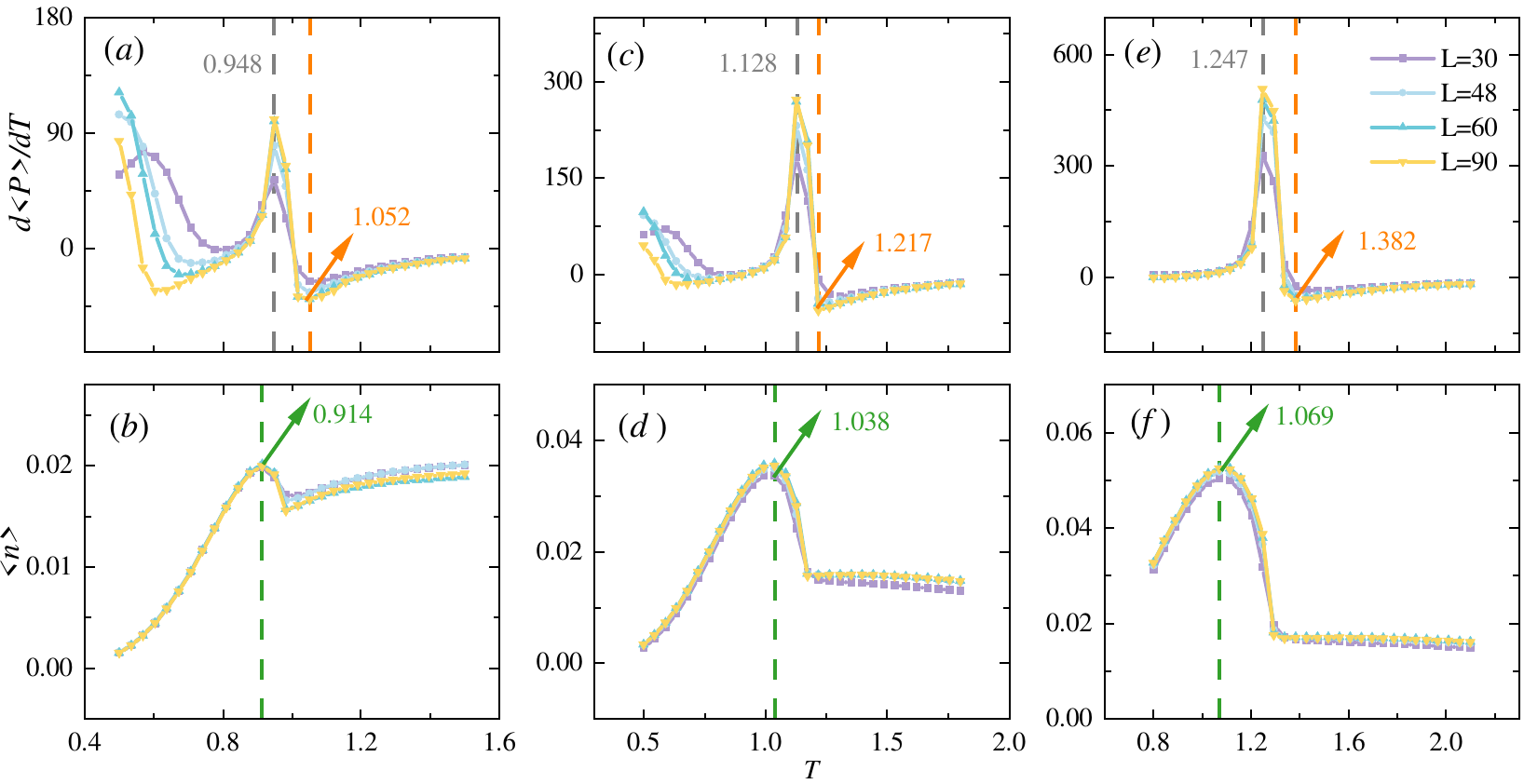}
\caption{Topology dependence of third-order precursor transitions in the three-state Potts model on Watts--Strogatz small-world networks. Panels (a),(b), (c),(d), and (e),(f) correspond to rewiring probabilities $p=0.1$, $0.3$, and $1.0$, respectively. The top row shows the perimeter-based post-critical signal $\mathrm{d}\langle P\rangle/\mathrm{d}T$, and the bottom row shows the isolated-spin indicator $\langle n_{\mathrm{iso}}\rangle$. Compared with the Ising case, the curves are smoother because multistate domain evolution is more gradual, but the same upward shift of $T_{\mathrm{ind}}$, $T_c$, and $T_{\mathrm{dep}}$ is preserved as rewiring increases. Gray dashed lines mark $T_c$, blue dashed lines mark $T_{\mathrm{ind}}$, and pink dashed lines mark $T_{\mathrm{dep}}$.}
\label{fig:potts_sw}
\end{figure}

The three-state Potts model displays the same overall temperature hierarchy on small-world networks, but with a smoother structural response and a stronger emphasis on boundary fluctuations. Fig.~\ref{fig:potts_sw} again uses the left, middle, and right columns for $p=0.1$, $0.3$, and $1.0$. As in the Ising case, increasing rewiring shifts all characteristic temperatures upward. Here the dominant peak of $\mathrm d\langle P\rangle/\mathrm dT$ is taken as the operational estimate of $T_c$, which rises from about $T_c\approx 0.95$ at $p=0.1$ to $\approx 1.13$ at $p=0.3$ and $\approx 1.25$ at $p=1.0$, while the post-critical precursor moves from $T_{\mathrm{dep}}\approx 1.05$ to $\approx 1.22$ and then to $\approx 1.38$.

The main difference from the Ising model lies in the structural channel through which the dependent transition is most visible. Because the Potts model supports three spin states, domain walls are more irregular and fluctuate more strongly even away from the critical point. As a result, $\mathrm{d}\langle P\rangle/\mathrm{d}T$ remains the most informative observable for the dependent precursor, whereas bulk area-based quantities are less discriminating. This is also why the Potts curves are smoother overall: multistate fragmentation distributes the structural rearrangement over a broader temperature interval than in the two-state Ising case.

The low-temperature precursor remains clearly visible through the maximum of $\langle n_{\mathrm{iso}}\rangle$, located at about $T_{\mathrm{ind}}\approx 0.91$, $1.04$, and $1.07$ for $p=0.1$, $0.3$, and $1.0$, respectively. Thus, despite the smoother curve shapes, the Potts model preserves the same qualitative hierarchy as the Ising model. The topology dependence is therefore robust across both models, while the model dependence appears primarily in which geometric observable best resolves the post-critical transition.

For completeness, Appendix~\hyperref[app:B]{B} examines the auxiliary composite observable $W$. Its post-critical extremum occurs near, but not exactly at, the perimeter-based $T_{\mathrm{dep}}$, which confirms that the precursor phenomenon is robust while also showing that the precise numerical location retains some observable dependence.

\subsection{Summary of characteristic temperatures}
\label{subsec:summary}

The numerical results across system sizes and rewiring probabilities support three robust conclusions. First, the ordering
\begin{equation}
T_{\mathrm{ind}} < T_c < T_{\mathrm{dep}}
\label{eq:hierarchy_summary}
\end{equation}
holds throughout the parameter range studied, on both the regular lattice and the small-world networks. Second, increasing the rewiring probability shifts all three temperatures upward in both models. Third, the topology-induced shifts are larger than the residual finite-size drift over the system sizes shown in the main figures, especially for the dependent transition.

Fig.~\ref{fig:temp_summary} makes these trends especially transparent. In both the Ising and Potts models, the strongest response to rewiring is seen in $T_{\mathrm{dep}}$, showing that post-critical boundary reorganization is particularly sensitive to long-range shortcuts. By contrast, $T_{\mathrm{ind}}$ varies more weakly with system size and with rewiring, which is consistent with its interpretation as the onset of cluster breakup rather than the main topology-amplified reorganization. The critical temperature $T_c$ lies between these two structural markers at all sizes, providing an internally consistent ordering across both models.

These observations support a coherent physical picture. The low-temperature precursor marks the onset of large-cluster fragmentation and isolated-spin proliferation, the conventional critical point corresponds to the dominant collective restructuring, and the high-temperature precursor reflects a further stage of post-critical boundary reorganization. Small-world rewiring does not alter this sequence; instead, it amplifies its structural visibility. This topology-driven enhancement is the central result of the present work.

\section{Conclusion}
\label{sec:conclusion}

We investigated third-order precursor transitions in the two-dimensional Ising and three-state Potts models on regular lattices and Watts--Strogatz small-world networks. By calibrating the conventional critical temperature independently from Binder-cumulant crossings and susceptibility peaks in the Ising model, and by locating the Potts critical point from the dominant peak of $\mathrm d\langle P\rangle/\mathrm dT$, we found a robust hierarchy of characteristic temperatures, $T_{\mathrm{ind}}<T_c<T_{\mathrm{dep}}$, on both regular lattices and shortcut-rich networks.

The main physical conclusion is that network topology does not erase this third-order hierarchy. On the contrary, increasing rewiring shifts the full set of characteristic temperatures upward and, most importantly, enhances the visibility of the post-critical precursor associated with boundary reorganization. In this sense, small-world shortcuts act less like noise and more like an amplifier of structurally meaningful fluctuation channels. Compared with the Ising case, the Potts model shows smoother domain evolution but a stronger sensitivity of perimeter-based observables to multistate boundary fluctuations, which makes interface measures especially useful for tracking the dependent transition.

These results indicate that higher-order precursor transitions can remain well defined even when local lattice geometry is replaced by heterogeneous shortcut connectivity. More broadly, the present framework combines independent critical-point calibration in the Ising case with structural observables that track cluster breakup and boundary reshaping, while in the Potts small-world case the main critical peak and post-critical precursor are resolved from the same perimeter-based derivative. Together these strategies provide a practical route to diagnosing subtle near-critical reorganizations in networked many-body systems.

\FloatBarrier
\phantomsection
\addcontentsline{toc}{chapter}{Appendix A: Additional finite-size scaling details}
\section*{Appendix A: Additional finite-size scaling details}
\label{app:A}

In the main text, finite-size analysis is used only to calibrate the conventional critical temperature $T_c$ for the Ising model independently of the cluster-based precursor observables. In the Potts model on small-world networks, the dominant peak of $\mathrm d\langle P\rangle/\mathrm dT$ is used operationally to locate $T_c$, while the subsequent post-critical extremum of the same curve identifies $T_{\mathrm{dep}}$. For completeness, we summarize here the role of the magnetization, susceptibility, and Binder-cumulant analysis in fixing the Ising critical region on Watts--Strogatz small-world networks.

The finite-size indicators discussed here should be interpreted as auxiliary consistency checks rather than as precision estimates of critical exponents. Because the focus of the present work is the identification of the precursor temperatures $T_{\mathrm{ind}}$ and $T_{\mathrm{dep}}$, we do not use this finite-size analysis as the central evidence for the main conclusions. Instead, its role is to provide an independent calibration of $T_c$ and to verify that the cluster-based precursor analysis is anchored to the conventional critical region.

As the system size increases, the Binder-cumulant crossings and susceptibility peaks become increasingly consistent, supporting the stability of the calibrated Ising critical temperature. The additional scaling information therefore reinforces the interpretation adopted in the main text while remaining secondary to the precursor analysis itself.

\FloatBarrier
\phantomsection
\addcontentsline{toc}{chapter}{Appendix B: Auxiliary structural observables in the Potts model}
\section*{Appendix B: Auxiliary structural observables in the Potts model}
\label{app:B}

In addition to the perimeter-based observable used in the main text, we also examined auxiliary composite quantities in the three-state Potts model in order to test the robustness of the post-critical structural transition. A representative example is the quantity
\begin{equation}
  W=\frac{1}{n_c}\sum_{C\in\mathcal{C}_2} A(C)\left[\frac{P(C)}{z\,A(C)}\right]^\alpha,
\end{equation}
where $z=4$ and $0<\alpha\le 1$.

\begin{figure}[htbp]
\centering
\includegraphics[width=\twobytwofig]{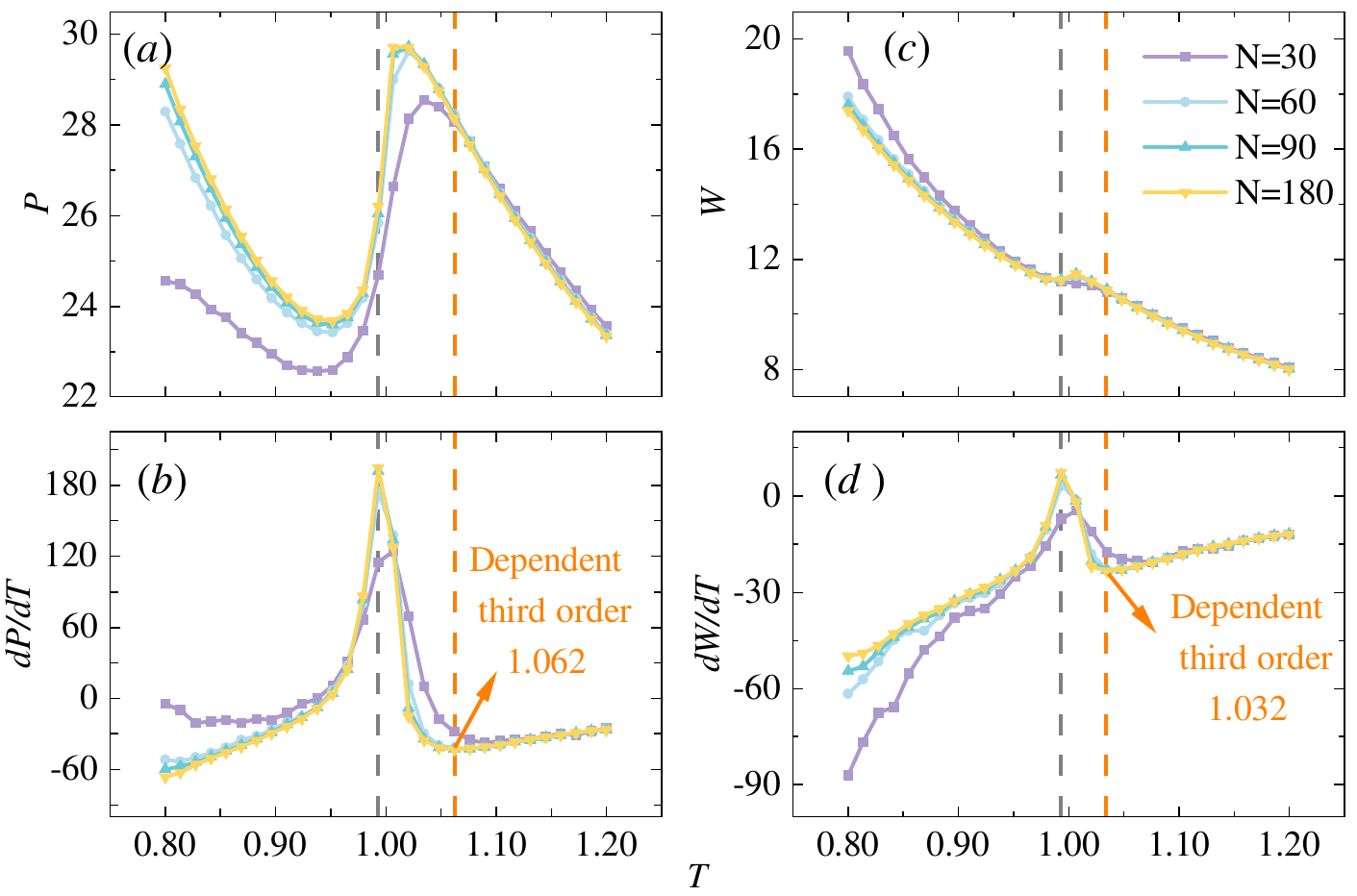}
\caption{Auxiliary composite observable $W$ in the three-state Potts model. Panels (a),(b) show $\langle P\rangle$ and $\mathrm{d}\langle P\rangle/\mathrm{d}T$; panels (c),(d) show $W$ and $\mathrm{d}W/\mathrm{d}T$. The derivative of $W$ exhibits a post-critical extremum close to, but not identical with, the perimeter-based $T_{\mathrm{dep}}$. This confirms the robustness of the post-critical reorganization while also showing that the precise numerical location of the precursor remains observable dependent.}
\label{fig:potts_W}
\end{figure}

Fig.~\ref{fig:potts_W} compares this auxiliary observable with the primary perimeter-based criterion. Panels~(a) and (b) show $\langle P\rangle$ and its temperature derivative, while panels~(c) and (d) show the composite observable $W$ and its derivative. The key point is that $\mathrm{d}W/\mathrm{d}T$ also develops a post-critical extremum, here near $T\approx 1.03$, whereas the perimeter-based criterion in the same data places the post-critical precursor near $T\approx 1.06$. The two observables therefore support the same qualitative interpretation---an additional structural reorganization on the disordered side of $T_c$---but they do not yield numerically identical precursor temperatures.

That small offset is physically reasonable. The quantity $W$ mixes cluster mass and boundary information, whereas $\mathrm{d}\langle P\rangle/\mathrm{d}T$ is more directly focused on interface restructuring. For this reason, we use the perimeter derivative as the primary operational definition of $T_{\mathrm{dep}}$ in the main text and regard $W$ as a robustness check rather than as an alternative primary criterion.

\FloatBarrier
\phantomsection
\addcontentsline{toc}{chapter}{Appendix C: Supplementary summary plots}
\section*{Appendix C: Supplementary summary plots}
\label{app:C}

Appendix~C collects compact summary plots that clarify how the characteristic temperatures and the corresponding signal strengths vary with system size and rewiring probability.

\begin{figure}[htbp]
\centering
\includegraphics[width=15.0cm]{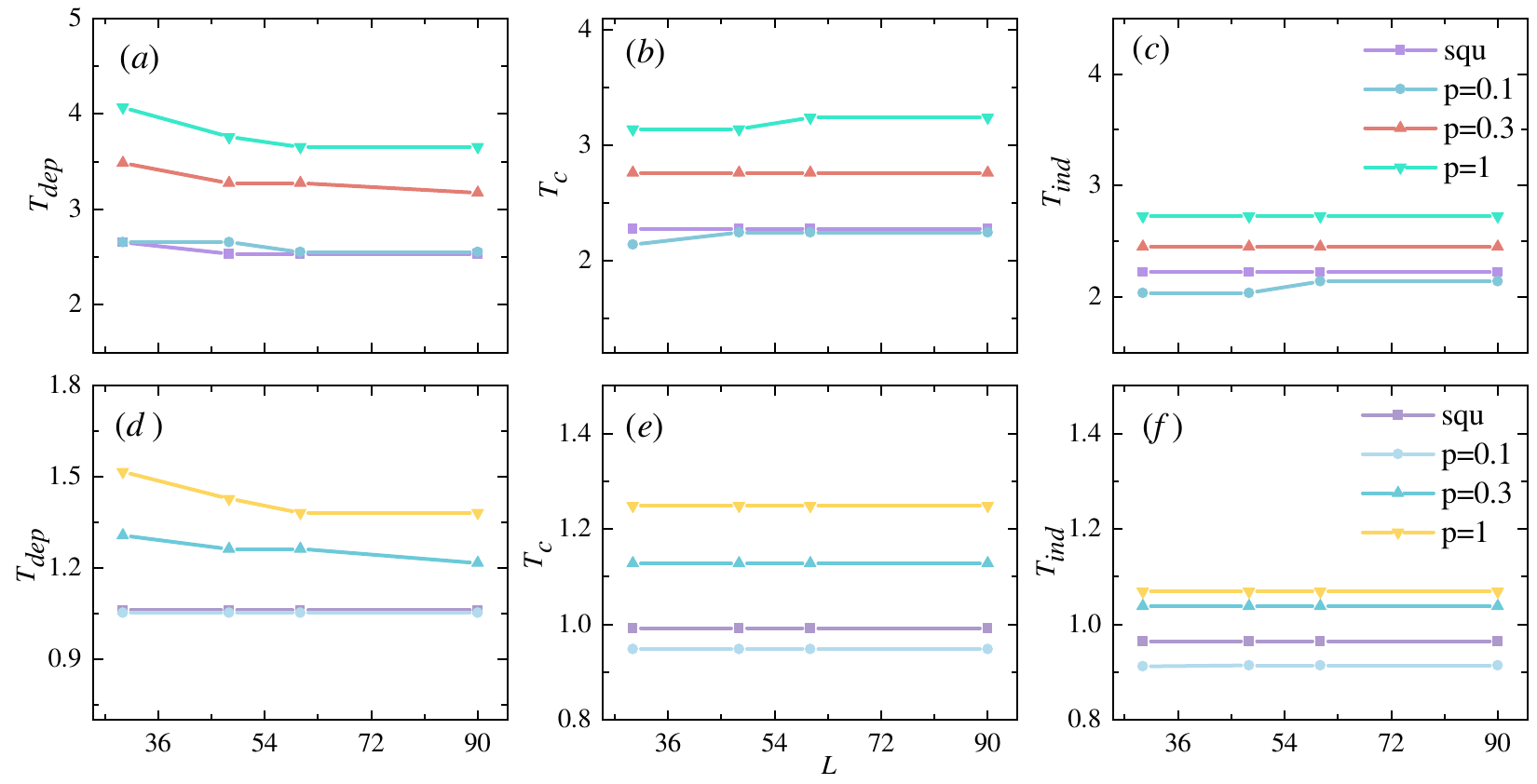}
\caption{Characteristic temperatures of the Ising and three-state Potts models on regular lattices and Watts--Strogatz small-world networks. Panels (a),(d) show the dependent third-order temperature $T_{\mathrm{dep}}$, panels (b),(e) the conventional critical temperature $T_c$, and panels (c),(f) the independent third-order temperature $T_{\mathrm{ind}}$. The top row corresponds to the Ising model and the bottom row to the Potts model. Across all system sizes and rewiring probabilities shown, the ordering $T_{\mathrm{ind}}<T_c<T_{\mathrm{dep}}$ is preserved, while increasing rewiring shifts the entire hierarchy upward, most strongly for the post-critical precursor.}
\label{fig:temp_summary}
\end{figure}

Fig.~\ref{fig:temp_summary} summarizes the three characteristic temperatures for both models. Panels~(a),(d) show $T_{\mathrm{dep}}$, panels~(b),(e) show $T_c$, and panels~(c),(f) show $T_{\mathrm{ind}}$, with the top row for the Ising model and the bottom row for the Potts model. Two features are especially clear. First, the ordering $T_{\mathrm{ind}}<T_c<T_{\mathrm{dep}}$ is preserved across all system sizes and network topologies studied here. Second, increasing rewiring shifts the entire hierarchy upward in both models, with the largest topology dependence appearing in $T_{\mathrm{dep}}$.

\begin{figure}[htbp]
\centering
\includegraphics[width=15.0cm]{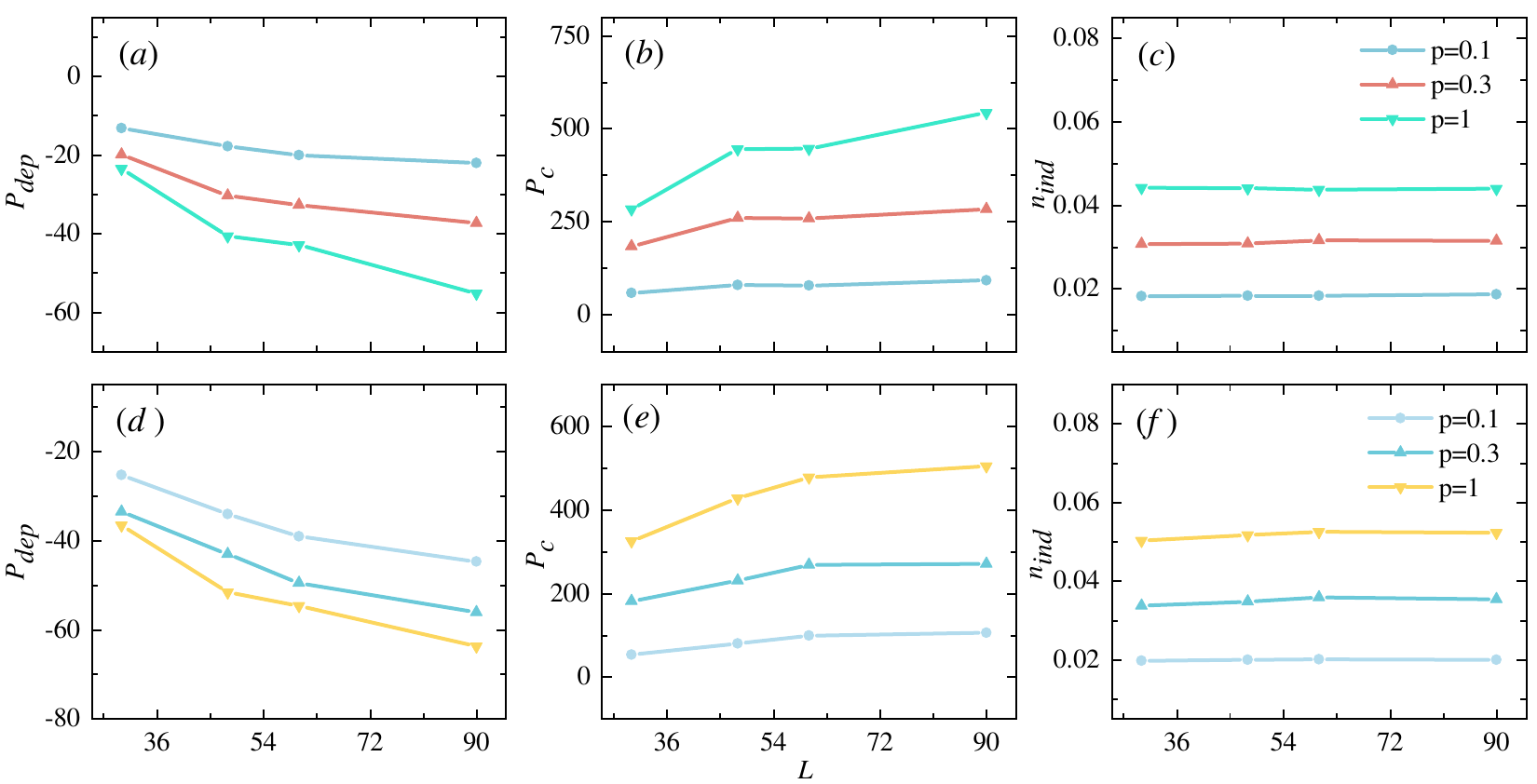}
\caption{Transition strengths of the precursor and critical features as functions of system size and rewiring probability for the Ising and three-state Potts models. Panels (a),(d) summarize the magnitude of the post-critical structural feature, panels (b),(e) the critical strength, and panels (c),(f) the low-temperature isolated-spin precursor. In both models, larger systems and stronger rewiring enhance the visibility of the structural signatures, with the strongest amplification occurring for the post-critical feature.}
\label{fig:strength_summary}
\end{figure}

Fig.~\ref{fig:strength_summary} summarizes the corresponding signal strengths. In the Ising case, both the critical strength $P_c$ and the magnitude of the post-critical extremum increase substantially with rewiring probability, while the independent-transition transition $n_{\mathrm{ind}}$ grows more moderately. The same qualitative trend is present in the Potts model. Thus, larger systems and stronger long-range rewiring not only shift the characteristic temperatures, but also improve the visibility of the precursor signatures themselves. These summary plots therefore reinforce the main conclusion that shortcut-enhanced connectivity amplifies, rather than obscures, the structural detectability of third-order precursor transitions.

\FloatBarrier

\FloatBarrier
\phantomsection
\addcontentsline{toc}{chapter}{Acknowledgment}
\section*{Acknowledgments}
\label{sec:ack}

This work was supported by China's National Natural Science Foundation Grant Nos.~12304257, 12322501, 12575035 and 12575033.
Y.T. acknowledges support from the Natural Science Foundation of Sichuan Province (Grant No.~2026NSFSCZY0124).
Computing resources were provided by the Interdisciplinary Intelligence Supercomputing Center of Beijing Normal University, Zhuhai.

\phantomsection
\addcontentsline{toc}{chapter}{References}
\bibliographystyle{iopart-num}
\bibliography{maincpb}

@PREAMBLE{
 "\providecommand{\noopsort}[1]{}"
 # "\providecommand{\singleletter}[1]{#1}%"
}

@article{Landau1999PhaseTransitionsCiSE,
  author  = {Landau, David P.},
  title   = {Phase Transitions and Critical Phenomena},
  journal = {Computing in Science \& Engineering},
  year    = {1999},
  volume  = {1},
  number  = {5},
  pages   = {10--11},
  doi     = {10.1109/MCSE.1999.790580}
}

@article{Ehrenfest1933Amsterdam,
  author  = {Ehrenfest, Paul},
  title   = {Phasenumwandlungen im ueblichen und erweiterten Sinn, classifiziert nach den entsprechenden Singularitaeten des thermodynamischen Potentiales},
  journal = {Proceedings of the Royal Academy of Amsterdam},
  year    = {1933},
  volume  = {36},
  pages   = {153}
}

@article{Fisher1972FiniteSizeScalingPRL,
  author  = {Fisher, Michael E. and Barber, Michael N.},
  title   = {Scaling Theory for Finite-Size Effects in the Critical Region},
  journal = {Physical Review Letters},
  year    = {1972},
  volume  = {28},
  number  = {23},
  pages   = {1516},
  doi     = {10.1103/PhysRevLett.28.1516}
}

@book{Gross2001MicrocanonicalThermodynamics,
  author    = {Gross, D. H. E.},
  title     = {Microcanonical Thermodynamics: Phase Transitions in ``Small'' Systems},
  publisher = {World Scientific},
  address   = {Singapore},
  year      = {2001}
}

@article{Qi2018MIPAClassification,
  author  = {Qi, Kai and Bachmann, Michael},
  title   = {Classification of Phase Transitions by Microcanonical Inflection-Point Analysis},
  journal = {Physical Review Letters},
  year    = {2018},
  volume  = {120},
  number  = {18},
  pages   = {180601},
  doi     = {10.1103/PhysRevLett.120.180601}
}

@article{Stevenson1981OPT,
  author  = {Stevenson, P. M.},
  title   = {Optimized perturbation theory},
  journal = {Physical Review D},
  volume  = {23},
  number  = {12},
  pages   = {2916},
  year    = {1981},
  doi     = {10.1103/PhysRevD.23.2916}
}

@article{Stevenson1981RSA,
  author  = {Stevenson, P. M.},
  title   = {Resolution of the renormalisation-scheme ambiguity in perturbative {QCD}},
  journal = {Physics Letters B},
  volume  = {100},
  number  = {1},
  pages   = {61--64},
  year    = {1981},
  doi     = {10.1016/0370-2693(81)90287-2}
}

@article{liu2022pseudo,
  author  = {Liu, Wei and Wang, Fangfang and Sun, Pengwei and Wang, Jincheng},
  title   = {Pseudo-Phase Transitions of {I}sing and {B}axter--{W}u Models in Two-Dimensional Finite-Size Lattices},
  journal = {Journal of Statistical Mechanics: Theory and Experiment},
  year    = {2022},
  volume  = {2022},
  number  = {9},
  pages   = {093206},
  doi     = {10.1088/1742-5468/ac8e5a}
}

@article{wangff,
  author  = {Wang, Fangfang and Liu, Wei and Ma, Jun and Qi, Kai and others},
  title   = {Exploring Transitions in Finite-Size {P}otts Model: Comparative Analysis Using {W}ang--{L}andau Sampling and Parallel Tempering},
  journal = {Journal of Statistical Mechanics: Theory and Experiment},
  year    = {2024},
  volume  = {2024},
  number  = {9},
  pages   = {093201},
  doi     = {10.1088/1742-5468/ad72da}
}

@article{BelHadjAissa2021KT,
  author  = {Bel-Hadj-Aissa, Ghofrane and Gori, Matteo and Franzosi, Roberto and Pettini, Marco},
  title   = {Geometrical and Topological Study of the {K}osterlitz--{T}houless Phase Transition in the {XY} Model in Two Dimensions},
  journal = {Journal of Statistical Mechanics: Theory and Experiment},
  year    = {2021},
  volume  = {2021},
  number  = {2},
  pages   = {023206},
  doi     = {10.1088/1742-5468/abda27}
}

@article{sitarachu2022evidence,
  author  = {Sitarachu, Kedkanok and Bachmann, Michael},
  title   = {Evidence for Additional Third-Order Transitions in the Two-Dimensional {I}sing Model},
  journal = {Physical Review E},
  year    = {2022},
  volume  = {106},
  number  = {1},
  pages   = {014134},
  doi     = {10.1103/PhysRevE.106.014134}
}

@article{Aierken2023,
  author  = {Aierken, Dilimulati and Bachmann, Michael},
  title   = {Stable Intermediate Phase of Secondary Structures for Semiflexible Polymers},
  journal = {Physical Review E},
  year    = {2023},
  volume  = {107},
  number  = {3},
  pages   = {L032501},
  doi     = {10.1103/PhysRevE.107.L032501}
}

@article{Aierken2023Bifurcation,
  author  = {Aierken, Dilimulati and Bachmann, Michael},
  title   = {Secondary-Structure Phase Formation for Semiflexible Polymers by Bifurcation in Hyperphase Space},
  journal = {Physical Chemistry Chemical Physics},
  year    = {2023},
  volume  = {25},
  number  = {44},
  pages   = {30246--30258},
  doi     = {10.1039/D3CP02815A}
}

@article{Corti2023,
  author  = {Corti, David S. and Ohadi, Donya and Fariello, Ricardo and Uline, Mark J.},
  title   = {Microcanonical Thermodynamics of Small Ideal Gas Systems},
  journal = {The Journal of Physical Chemistry B},
  year    = {2023},
  volume  = {127},
  number  = {15},
  pages   = {3431--3442},
  doi     = {10.1021/acs.jpcb.3c00278}
}

@article{DiCairano2024Topological,
  author  = {Di Cairano, Loris and Gori, Matteo and Sarkis, Matthieu and Tkatchenko, Alexandre},
  title   = {Detecting Phase Transitions in Lattice Gauge Theories: Production and Dissolution of Topological Defects in {4D} Compact Electrodynamics},
  journal = {Physical Review D},
  year    = {2024},
  volume  = {110},
  number  = {1},
  pages   = {014503},
  doi     = {10.1103/PhysRevD.110.014503}
}

@article{DiCairano2022Topological,
  author  = {Di Cairano, Loris and Capelli, Riccardo and Bel-Hadj-Aissa, Ghofrane and Pettini, Marco},
  title   = {Topological Origin of the Protein Folding Transition},
  journal = {Physical Review E},
  year    = {2022},
  volume  = {106},
  number  = {5},
  pages   = {054134},
  doi     = {10.1103/PhysRevE.106.054134}
}

@article{shi2026pseudo,
  author  = {Shi, Lei and Liu, Wei and Qi, Kai and Xiong, Kezhao and others},
  title   = {Pseudo transitions in the finite-size six-state clock model},
  journal = {Communications in Theoretical Physics},
  volume  = {78},
  number  = {3},
  pages   = {035601},
  year    = {2026},
  doi     = {10.1088/1572-9494/ae1188}
}

@article{liu2025PLA,
  author  = {Liu, Wei and Shi, Lei and Zhang, Xin and Li, Xiang and others},
  title   = {Pseudo transitions in the finite-size {B}lume--{C}apel model},
  journal = {Physics Letters A},
  volume  = {552},
  pages   = {130626},
  year    = {2025},
  doi     = {10.1016/j.physleta.2025.130626}
}

@article{Liu2024MV,
  author  = {Liu, Wei and Wang, Jincheng and Wang, Fangfang and Qi, Kai and others},
  title   = {The precursor of the critical transitions in majority vote model with the noise feedback from the vote layer},
  journal = {Journal of Statistical Mechanics: Theory and Experiment},
  year    = {2024},
  volume  = {2024},
  number  = {8},
  pages   = {083402},
  doi     = {10.1088/1742-5468/ad6426}
}

@article{Rocha2025Microcanonical,
  author  = {Rocha, J. C. S. and Dias, R. A. and Costa, B. V.},
  title   = {Microcanonical inflection-point analysis via parametric curves and its relation to the zeros of the partition function},
  journal = {Physical Review E},
  volume  = {112},
  number  = {1},
  pages   = {014112},
  year    = {2025},
  doi     = {10.1103/PhysRevE.112.014112}
}

@article{Junghans2006PeptideAggregation,
  author  = {Junghans, Christoph and Bachmann, Michael and Janke, Wolfhard},
  title   = {Microcanonical Analyses of Peptide Aggregation Processes},
  journal = {Physical Review Letters},
  volume  = {97},
  number  = {21},
  pages   = {218103},
  year    = {2006},
  doi     = {10.1103/PhysRevLett.97.218103}
}

@article{Schnabel2011EntropyInflection,
  author  = {Schnabel, Stefan and Seaton, Daniel T. and Landau, David P. and Bachmann, Michael},
  title   = {Microcanonical entropy inflection points: Key to systematic understanding of transitions in finite systems},
  journal = {Physical Review E},
  volume  = {84},
  number  = {1},
  pages   = {011127},
  year    = {2011},
  doi     = {10.1103/PhysRevE.84.011127}
}

@article{Wang2026Canonical,
  author  = {Wang, Fangfang and Liu, Wei and Qi, Kai and Cui, Zidong and others},
  title   = {Canonical Criterion for Third-Order Transitions},
  journal = {arXiv preprint arXiv:2603.09124},
  year    = {2026},
  doi     = {10.48550/arXiv.2603.09124}
}

@article{Liu2025PottsGeometry,
  author  = {Liu, Wei and Zhang, Xin and Shi, Lei and Qi, Kai and others},
  title   = {Geometric properties of the additional third-order transitions in the two-dimensional {P}otts model},
  journal = {Physical Review E},
  volume  = {111},
  number  = {5},
  pages   = {054128},
  year    = {2025},
  doi     = {10.1103/PhysRevE.111.054128}
}

@article{Sitarachu2020ThirdOrderIsing,
  author  = {Sitarachu, Kedkanok and Bachmann, Michael},
  title   = {Third-Order Phase Transitions in the Two-Dimensional {I}sing Model},
  journal = {Journal of Physics: Conference Series},
  year    = {2020},
  volume  = {1483},
  number  = {1},
  pages   = {012009},
  doi     = {10.1088/1742-6596/1483/1/012009}
}

@article{SWN,
  author  = {Watts, Duncan J. and Strogatz, Steven H.},
  title   = {Collective dynamics of `small-world' networks},
  journal = {Nature},
  volume  = {393},
  number  = {6684},
  pages   = {440--442},
  year    = {1998},
  doi     = {10.1038/30918}
}

@article{AlbertBarabasi2002RMP,
  author  = {Albert, R{\'e}ka and Barab{\'a}si, Albert-L{\'a}szl{\'o}},
  title   = {Statistical mechanics of complex networks},
  journal = {Reviews of Modern Physics},
  volume  = {74},
  number  = {1},
  pages   = {47--97},
  year    = {2002},
  doi     = {10.1103/RevModPhys.74.47}
}

@article{Dorogovtsev2008CriticalPhenomenaNetworks,
  author  = {Dorogovtsev, S. N. and Goltsev, A. V. and Mendes, J. F. F.},
  title   = {Critical phenomena in complex networks},
  journal = {Reviews of Modern Physics},
  volume  = {80},
  number  = {4},
  pages   = {1275--1335},
  year    = {2008},
  doi     = {10.1103/RevModPhys.80.1275}
}

@article{Gitterman2000SmallWorldIsing,
  author  = {Gitterman, Moshe},
  title   = {Small-world phenomena in physics: The {I}sing model},
  journal = {Journal of Physics A: Mathematical and General},
  volume  = {33},
  number  = {47},
  pages   = {8373--8381},
  year    = {2000},
  doi     = {10.1088/0305-4470/33/47/101}
}

@article{Herrero2002IsingSmallWorld,
  author  = {Herrero, Carlos P.},
  title   = {Ising model in small-world networks},
  journal = {Physical Review E},
  volume  = {65},
  number  = {6},
  pages   = {066110},
  year    = {2002},
  doi     = {10.1103/PhysRevE.65.066110}
}

@article{Herrero2004FerroSmallWorld,
  author  = {Herrero, Carlos P.},
  title   = {Finite-size scaling analysis of the {I}sing model on small-world networks},
  journal = {Physical Review E},
  volume  = {69},
  number  = {6},
  pages   = {067109},
  year    = {2004},
  doi     = {10.1103/PhysRevE.69.067109}
}

@article{Ferrenberg1988MonteCarlo,
  author  = {Ferrenberg, Alan M. and Swendsen, Robert H.},
  title   = {New Monte Carlo Technique for Studying Phase Transitions},
  journal = {Physical Review Letters},
  volume  = {61},
  number  = {23},
  pages   = {2635--2638},
  year    = {1988},
  doi     = {10.1103/PhysRevLett.61.2635}
}

@article{Swendsen1987Nonuniversal,
  author  = {Swendsen, Robert H. and Wang, Jian-Sheng},
  title   = {Nonuniversal critical dynamics in Monte Carlo simulations},
  journal = {Physical Review Letters},
  volume  = {58},
  number  = {2},
  pages   = {86--88},
  year    = {1987},
  doi     = {10.1103/PhysRevLett.58.86}
}

@article{ising,
  author  = {Ising, Ernst},
  title   = {Beitrag zur Theorie des Ferromagnetismus},
  journal = {Zeitschrift f{\"u}r Physik},
  year    = {1925},
  volume  = {31},
  pages   = {253--258},
  doi     = {10.1007/BF02980577}
}

@article{WuRMP1982,
  author  = {Wu, F. Y.},
  title   = {The {P}otts Model},
  journal = {Reviews of Modern Physics},
  year    = {1982},
  volume  = {54},
  number  = {1},
  pages   = {235--268},
  doi     = {10.1103/RevModPhys.54.235}
}

@article{Potts1952,
  author  = {Potts, Renfrey B.},
  title   = {Some Generalized Order-Disorder Transformations},
  journal = {Proceedings of the Cambridge Philosophical Society},
  year    = {1952},
  volume  = {48},
  number  = {1},
  pages   = {106--109},
  doi     = {10.1017/S0305004100027419}
}

\end{CJK*}
\end{document}